\documentclass{aa}  

\usepackage{graphicx}
\usepackage{txfonts}
\usepackage{color}
\begin{document}

   \title{Modelling circumbinary protoplanetary disks}

   \subtitle{I. Fluid simulations of the Kepler-16 and 34 systems.}

   \author{S. Lines
          \inst{1}
          \and
          Z. M. Leinhardt\inst{1}\fnmsep
          \and
          C. Baruteau\inst{2,3}\fnmsep
          \and
          S.-J. Paardekooper\inst{4,5}\fnmsep
          \and
          P. J. Carter\inst{1}\fnmsep
          }

   \institute{School of Physics, University of Bristol, H. H. Wills Physics Laboratory, Tyndall Avenue, Bristol, BS8 1TL, UK\\
              \email{stefan.lines@bristol.ac.uk, zoe.leinhardt@bristol.ac.uk, p.carter@bristol.ac.uk}
         \and
             CNRS, IRAP, 14 avenue Edouard Belin, 31400 Toulouse, France
          \and
             Universit{\'e} de Toulouse, UPS-OMP, IRAP, Toulouse, France\\
             \email{clement.baruteau@irap.omp.eu }
          \and
             Astronomy Unit, School of Physics $\&$ Astronomy, Queen Mary University of London, UK\\
             \email{s.j.paardekooper@qmul.ac.uk}
          \and
             DAMTP, University of Cambridge, Wilberforce Road, Cambridge CB3 0WA, UK
             }

   \date{Accepted July 21, 2015}

% \abstract{}{}{}{}{} 
% 5 {} token are mandatory
 
  \abstract
   {The Kepler mission's discovery of a number of circumbinary planets orbiting close ($a_p$ $<$ 1.1 au) to the stellar binary raises questions as to how these planets could have formed given the intense gravitational perturbations the dual stars impart on the disk. The gas component of circumbinary protoplanetary disks is perturbed in a similar manner to the solid, planetesimal dominated counterpart, although the mechanism by which disk eccentricity originates differs.}
   {This is the first work of a series that aims to investigate the conditions for planet formation in circumbinary protoplanetary disks.}
   {We present a number of hydrodynamical simulations that explore the response of gas disks around two observed binary systems: Kepler-16 and Kepler-34. We probe the importance of disk viscosity, aspect-ratio, inner boundary condition, initial surface density gradient, and self-gravity on the dynamical evolution of the disk, as well as its quasi-steady-state profile.}
   {We find there is a strong influence of binary type on the mean disk eccentricity, $\bar{e}_d$, leading to $\bar{e}_d = 0.02 - 0.08$ for Kepler-16 and $\bar{e}_d = 0.10 - 0.15$ in Kepler-34. The value of $\alpha$-viscosity has little influence on the disk, but we find a strong increase in mean disk eccentricity with increasing aspect-ratio due to wave propagation effects. The choice of inner boundary condition only has a small effect on the surface density and eccentricity of the disk. Our primary finding is that including disk self-gravity has little impact on the evolution or final state of the disk for disks with masses less than 12.5 times that of the minimum-mass solar nebula. This finding contrasts with the results of self-gravity relevance in circumprimary disks, where its inclusion is found to be an important factor in describing the disk evolution.}
   {}


   \maketitle
%
%________________________________________________________________

\section{Introduction}

Extrasolar planets in circumbinary (p-type) orbits around short-period stellar binary systems are a recent addition to the increasingly diverse range of planetary characteristics found outside our solar system. We now know of nine short-period planets\footnote{Kepler-16(AB)b \citep{doyle11}, Kepler-34(AB)b $\&$ 35(AB)b \citep{welsh12}, Kepler-38(AB)b \citep{orosz12a}, Kepler-47(AB)b,c \citep{orosz12b}, Kepler-64(AB)b \citep{schwamb13}, Kepler-413(AB)b \citep{kostov14} and KIC 9632895 \citep{welsh14}.} with semi-major axis $a_p <$ 1.1 au, placing them close to the binary barycentre and under the influence of perturbative forces from the secondary star.

The stability of circumbinary planet orbits was studied well before the first confirmed exoplanet detection in 1992 \citep{wolszczan92} by \cite{heppenheimer78}. More recently, \cite{holman99} performed a parameter space study of the long-term stability of planets in P-type orbits. They calculated a range of orbital radii that allows for dynamical stability for a range of binary eccentricities and semi-major axes. Interestingly, with the exception of Kepler-47(AB)c, all known circumbinary planets lie just outside their inner most stable orbit, $a_c$. One plausible explanation for this occurrence is that at some point in their evolution, these planets underwent a period of migration that was halted. 

\cite{pierens08b} showed that Saturn- and Jupiter-mass planets undergo different interactions with the binary that can result in either stabilisation of the planetary orbits or its removal from the system via ejection or scattering. Specifically they find that after a period of inwards migration, Saturn-mass planets have a stable evolution. A torque reversal, caused by an eccentricity increase from interaction with the binary, leads to stable outwards migration. However, Jupiter-mass planets often become trapped in 4:1 mean motion resonance with the binary, which causes significant increases in planetary eccentricity \citep{nelson03}. Eventually, close encounters with the stars at periapsis lead to outward scattering or complete ejection. The latter effect may explain why all Kepler circumbinary planets discovered so far are sub-Jupiter mass, a result difficult to explain by observational biases since larger planets are easier to detect.

\begin{figure*}
\centering
\includegraphics[scale=1.2,natwidth=551,natheight=334]{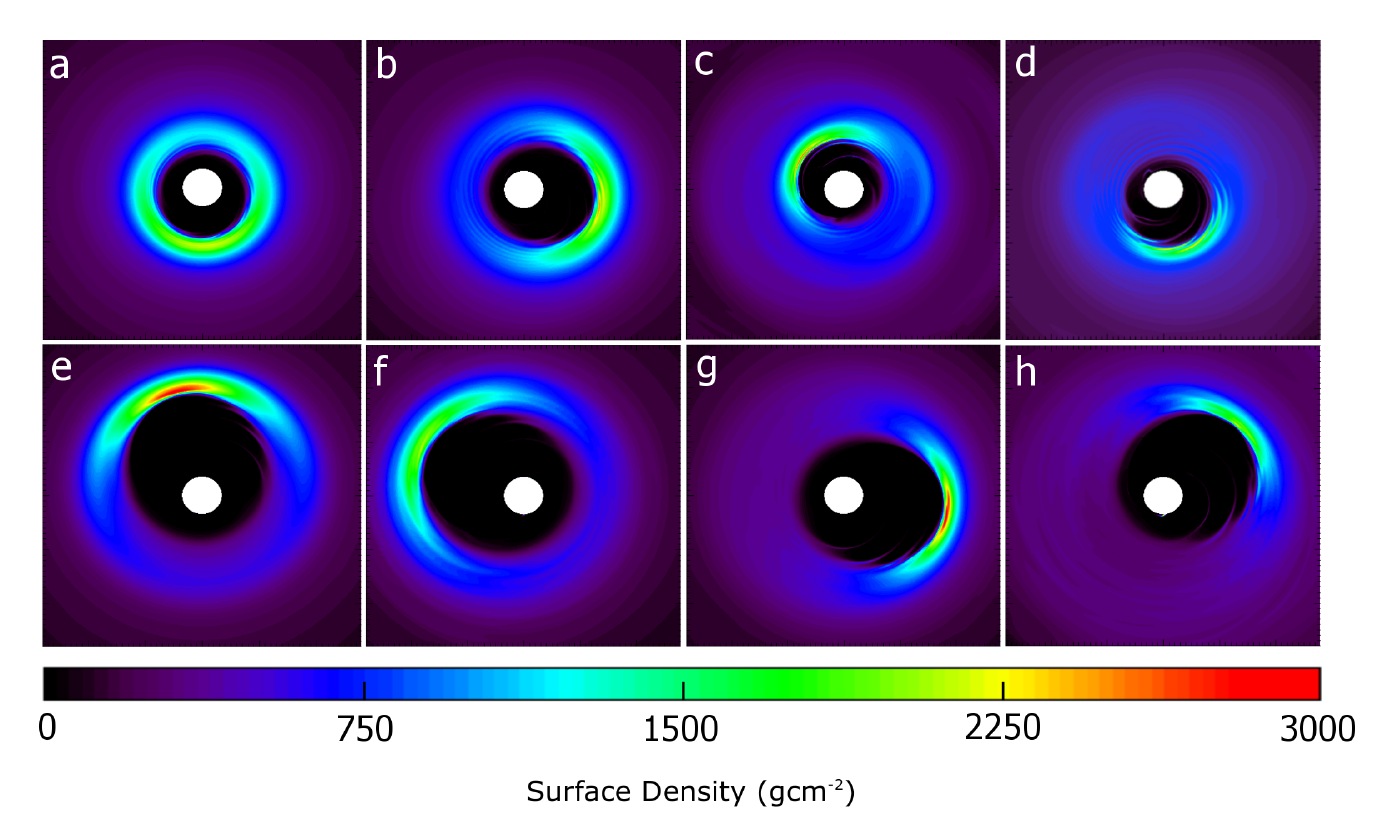}
\vspace{-10pt}
\caption{Surface density maps, $\Sigma_{2D}$, for non self-gravitating models around Kepler-16 (top row) and Kepler-34 (bottom row) at 16,000 $P_{AB}$. Frames a-d and e-h correspond to simulations $A$-$D$ and $E$-$H$ in Table \ref{tab:fargomodels}, respectively.
}
\vspace{+10pt}
\label{fig:panel}
\end{figure*}

\begin{table*}[t]
\centering
\vspace{+15pt}
\begin{tabular}{c|c|c|c|c|c|c|c}
\bf {Simulation} & \bf{Binary System} & \bf{$h$} & \bf{$\alpha$} & \bf{$\Sigma_0$ ($M_{\odot}/au^{2}$)} & \bf{Self-Gravity} & \bf{Boundary} & \bf{$\alpha_\Sigma$}  \\ \hline \hline
\bf{A} & Kepler-16 & 0.03 & $10^{-3}$ & $1.0\times10^{-4}$ & No & Rigid & 1.5\\
\bf{B} & Kepler-16 & 0.05 & $10^{-3}$ & $1.0\times10^{-4}$ & No & Rigid & 1.5\\
\bf{C} & Kepler-16 & 0.03 & $10^{-2}$ & $1.0\times10^{-4}$ & No & Rigid & 1.5\\
\bf{D} & Kepler-16 & 0.05 & $10^{-2}$ & $1.0\times10^{-4}$ & No & Rigid & 1.5\\
\bf{E} & Kepler-34 & 0.03 & $10^{-3}$ & $1.0\times10^{-4}$ & No & Rigid & 1.5\\
\bf{F} & Kepler-34 & 0.05 & $10^{-3}$ & $1.0\times10^{-4}$ & No & Rigid & 1.5\\
\bf{G} & Kepler-34 & 0.03 & $10^{-2}$ & $1.0\times10^{-4}$ & No & Rigid & 1.5\\
\bf{H} & Kepler-34 & 0.05 & $10^{-2}$ & $1.0\times10^{-4}$ & No & Rigid & 1.5\\ \hline
\bf{I} & Kepler-16 & 0.03 & $10^{-3}$ & $1.0\times10^{-4}$ & Yes & Rigid & 1.5\\
\bf{J} & Kepler-16 & 0.05 & $10^{-3}$ & $1.0\times10^{-4}$ & Yes & Rigid & 1.5\\
\bf{K} & Kepler-34 & 0.03 & $10^{-3}$ & $1.0\times10^{-4}$ & Yes & Rigid & 1.5\\
\bf{L} & Kepler-34 & 0.05 & $10^{-3}$ & $1.0\times10^{-4}$ & Yes & Rigid & 1.5\\ \hline
\bf{M} & Kepler-16 & 0.05 & $10^{-3}$ & $1.0\times10^{-4}$ & No & Open & 1.5\\
\bf{N} & Kepler-34 & 0.05 & $10^{-3}$ & $1.0\times10^{-4}$ & No & Open & 1.5\\ \hline
\bf{O} & Kepler-16 & 0.05 & $10^{-3}$ & $5.0\times10^{-4}$ & Yes & Rigid & 1.5\\
\bf{P} & Kepler-16 & 0.05 & $10^{-3}$ & $1.0\times10^{-3}$ & Yes & Rigid & 1.5\\
\bf{Q} & Kepler-16 & 0.05 & $10^{-3}$ & $2.5\times10^{-3}$ & Yes & Rigid & 1.5\\
\bf{R} & Kepler-16 & 0.05 & $10^{-3}$ & $5.0\times10^{-4}$ & No & Rigid & 1.5\\
\bf{S} & Kepler-16 & 0.05 & $10^{-3}$ & $1.0\times10^{-3}$ & No & Rigid & 1.5\\
\bf{T} & Kepler-16 & 0.05 & $10^{-3}$ & $2.5\times10^{-3}$ & No & Rigid & 1.5\\ \hline
\bf{U} & Kepler-16 & 0.05 & $10^{-3}$ & $1.0\times10^{-4}$ & Yes & Rigid & 0.5\\ \hline
\end{tabular}
\label{tab:fargomodels}
\vspace{+10pt}
\caption{Parameter setup of each simulation A through U. A-H test the effect of varying aspect ratio, alpha-viscosity and central binary type on non self-gravitating disks. I-L test the inclusion of self-gravity at standard surface density. M-N investigate the inclusion of an open boundary. O-T look at the relevance of self-gravity in more detail by increasing the surface density of the disk. U tests the dependency of the disk evolution on the surface density profile exponent, $\alpha_\Sigma$.}
\vspace{+10pt}
\end{table*}

Although all the observed circumbinary planets are currently in stable orbits, it is not obvious that they formed in situ. Several studies have been carried out to try to understand how the processes behind planet formation and evolution can occur in spite of an intense gravitational influence on the protoplanetary disk from the central binary. One of the key hurdles in forming solid cores and terrestrial planets in circumbinary disks is explaining how rocky, metre- to kilometre-sized planetesimals undergo collisions that result in growth rather than erosion. It has been shown in previous work that eccentricity forcing from the binary on the protoplanetary disk can drive up relative velocities which not only inhibits accretion but can lead to energetic and highly erosive impacts. In particular, \cite{lines14} showed that even massive planetesimals are often disrupted despite having a large gravitational binding energy. Their work corroborates that of \cite{paardekooper12} and \cite{meschiari12a,meschiari12b} suggesting that sustained planetesimal accretion is unlikely and thus, the observed Kepler circumbinary planets did not form in situ. The most likely explanation for the presence of these planets is the migration of planetary cores that form at larger orbital radii where the gravitational forcing from the binary is of significantly lower magnitude.

A key component missing from many of the \emph{N}-body studies of planetesimal evolution in circumbinary disks is the inclusion of a gas disk which is typically 100 times the mass of its solid counterpart. As is the case with the rocky planetesimals, the gas is also perturbed by the binary, raising the eccentricity of the disk. The eccentricity in the gas disk is established through parametric instabilities originating from non-linear mode coupling between the eccentric $m = 1$ modes of both the initial disk eccentricity and the binary potential \citep{papaloizou02,hayasaki09,lubow91} and well as the direct driving from the binary $m = 1$ potential \citep{lubow00}. This coupling produces an $m = 2$ spiral wave at the 3:1 Lindblad resonance which increases disk eccentricity via the outwards transport of angular momentum. The overall result is an eccentric, asymmetric disk precessing about the binary \citep[(hereafter referred to as PN13)]{kley14,marzari13,pelupessy13,pierens13}. The asymmetric gas disk interacts with the solid planetesimal disk through gas drag and the time dependent gravitational potential of the gas disk. The latter effect is particularly important when large asymmetries in the gas surface density arise (see Figure \ref{fig:panel}).

The role of gas drag has been probed in several studies which found that differential orbital phasing caused relative velocities to remain small between planetesimals of comparative mass, thus, orbit crossing will occur between planetesimals of different sizes \citep{scholl07,marzari08}.
The influence of disk self-gravity on the evolution of the system is less well explored. Recent studies such as PN13 and \cite{kley14} choose not to include self-gravity in their simulations. The former justifies this approximation by confirming the Toomre parameter,

\begin{equation}
Q = \frac{c_s \Omega}{\pi G \Sigma},
\end{equation}
where $c_s$ is the sound speed, $\Omega$ is the angular velocity, $G$ is the gravitational constant, and $\Sigma$ is the surface density, satisfies the Toomre stability criterion ($Q>1$) at all disk radii for their chosen disk mass. However, even in self-gravitating disks of relatively low mass, low-frequency global modes exist that do not have a counterpart in non self-gravitating disks \citep{papaloizou02}. Moreover, the study by \cite{marzari09} shows that self-gravity in circumprimary disks can be important even in low-mass disks. Therefore, an investigation into whether or not self-gravity is required to fully describe a circumbinary disk, even at lower masses, is required.

In this paper, we aim to explore a number of parameters characterising a gaseous circumbinary disk in an attempt to find a suitable surface density profile for both the Kepler-16 and Kepler-34 systems. The resulting quasi-steady-state surface density profiles will be used to parameterize the gas disk potential for use in our $N$-body simulations of planetesimal growth. This next work will be presented in a second paper, part II. Section \ref{sec:method} presents the numerical method and model parameters. In Section \ref{sec:results} and \ref{sec:dis} we discuss the results and implications of the 21 simulations of the Kepler-16 and Kepler-34 systems. Finally, we summarise our findings in the conclusion in Section \ref{sec:summary}.

%__________________________________________________________________

\section{Numerical methods}\label{sec:method}

%                                     Two column figure (place early!)
%______________________________________________ Gamma_1 (lg rho, lg e)
   The central stellar binary potential is calculated in a fixed steady state orbit centred on the binary barycentre, meaning that the stars do not respond to the gas disk \citep{pierens07}. The stars, with mass ratio $q=M_B/M_A$ are first initialised at their respective apoapses, such that the Cartesian coordinates are initially zero in the $Y$ direction for both stars, $Y_A$ $\&$ $Y_B$ = 0, and the $X$ position is
\begin{equation}
X_A = -\frac{q}{1+q}r,
\end{equation}
and
\begin{equation}
X_B = \frac{1}{1+q}r,
\end{equation}
where $M_A$ and $M_B$ are the mass of star $A$ and $B$, respectively, and the distance of a star from its focus, r, is determined from the shape equation
\begin{equation}
r = a_b\left(\frac{1 - e_b^2}{1-e_b \cos \theta}\right),
\end{equation}
where $a_b$ is the binary semi-major axis, $e_b$ is the binary eccentricity, and the angle of the star from apoapsis is $\theta$.
Using Kepler's 2nd law to determine the fractional area swept out at any given time, an iterative procedure is invoked to calculate $\theta(t)$ and thus, the position vectors of each star. The stellar binary parameters are given in Table \ref{tab:binaryparams}.

\begin{figure*}
\centering
\includegraphics[scale=1.2,natwidth=550,natheight=224]{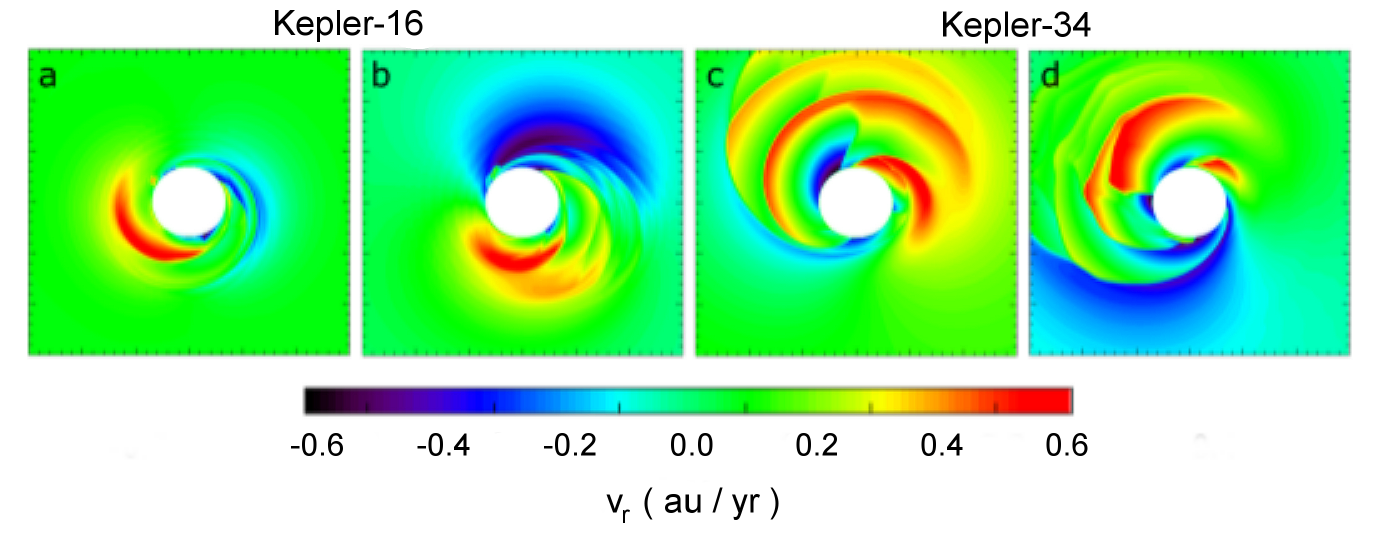}
\vspace{-10pt}
\caption{Radial velocity maps at 16,000 $P_{AB}$ for a) Kepler-16 $\alpha$ = 10$^{-3}$, $h$ = 0.03, b) Kepler-16 $\alpha$ = 10$^{-3}$, $h$ = 0.05, c) Kepler-34 $\alpha$ = 10$^{-3}$, $h$ = 0.03 and d) Kepler-34 $\alpha$ = 10$^{-3}$, $h$ = 0.05. }
\vspace{+10pt}
\label{fig:panel2}
\end{figure*}

To simulate the hydrodynamical evolution of the circumbinary disks we use a modified version of the Eulerian fluid code FARGO \citep{masset00}, called FARGO-ADSG \citep{baruteau08}, which includes optional disk self-gravity.
The hydrodynamical equations are solved on a polar mesh which is composed of $n_\phi$ = 512 equally divided azimuthal sector cells, and $n_r$ = 395 logarithmically spaced radial cells. Logarithmic spacing in the radial direction is a necessary condition for the fast fourier transform algorithm in the self-gravity calculation, but is also a preference for achieving the highest resolution closest to the binary \citep{baruteau08}.

The majority of our runs use a rigid, reflecting inner boundary condition which is fixed at 0.345 au with the rigid outer boundary at 4.0 au. In a couple of simulations we have used standard outflow boundaries at the grid's inner and outer edges, where the zero-gradient condition has been extended to the gas azimuthal velocity. This is a necessary implementation since there is no well defined equilibrium between the central gravity from the binary and the opposing centrifugal force and pressure gradient. In all simulations the inner disk edge is truncated by the binary causing a steep positive density gradient which is numerically difficult to model smoothly, thus, we employ a gap function, $f_{gap}$ from \citet{gunther04} to better model the initial surface density profile in the inner disk region:
\begin{equation}
f_{gap} = \left(1+\exp\left({-\frac{R-R_{gap}}{0.1R_{gap}}}\right)\right)^{-1},
\end{equation}
where $R_{gap} = 2.5a_b$ is the estimated size of the gap \citep{artymowicz94} and $R$ is the orbital radius.

The simulated gas disk is simplified by using the isothermal disk approximation, thereby minimising the number of unknowns. We leave the inclusion of an energy equation to account for the thermal evolution of the disk for a future paper. 

\begin{table}
\centering
\vspace{+15pt}
\begin{tabular}{c|c|c|c}
\bf {System} & \bf{$q$} & \bf{$e_b$} & {$a_b$ (au)}\\ \hline \hline
\bf{Kepler-16} & 0.29 & 0.16 & $0.22$\\
\bf{Kepler-34} & 1.0 & 0.52 & $0.23$\\ \hline
\end{tabular}
\label{tab:binaryparams}
\vspace{+10pt}
\caption{Stellar binary parameters.}
\end{table}

\begin{figure*}
\centering
\vspace{+0pt}
\includegraphics[scale=1.22]{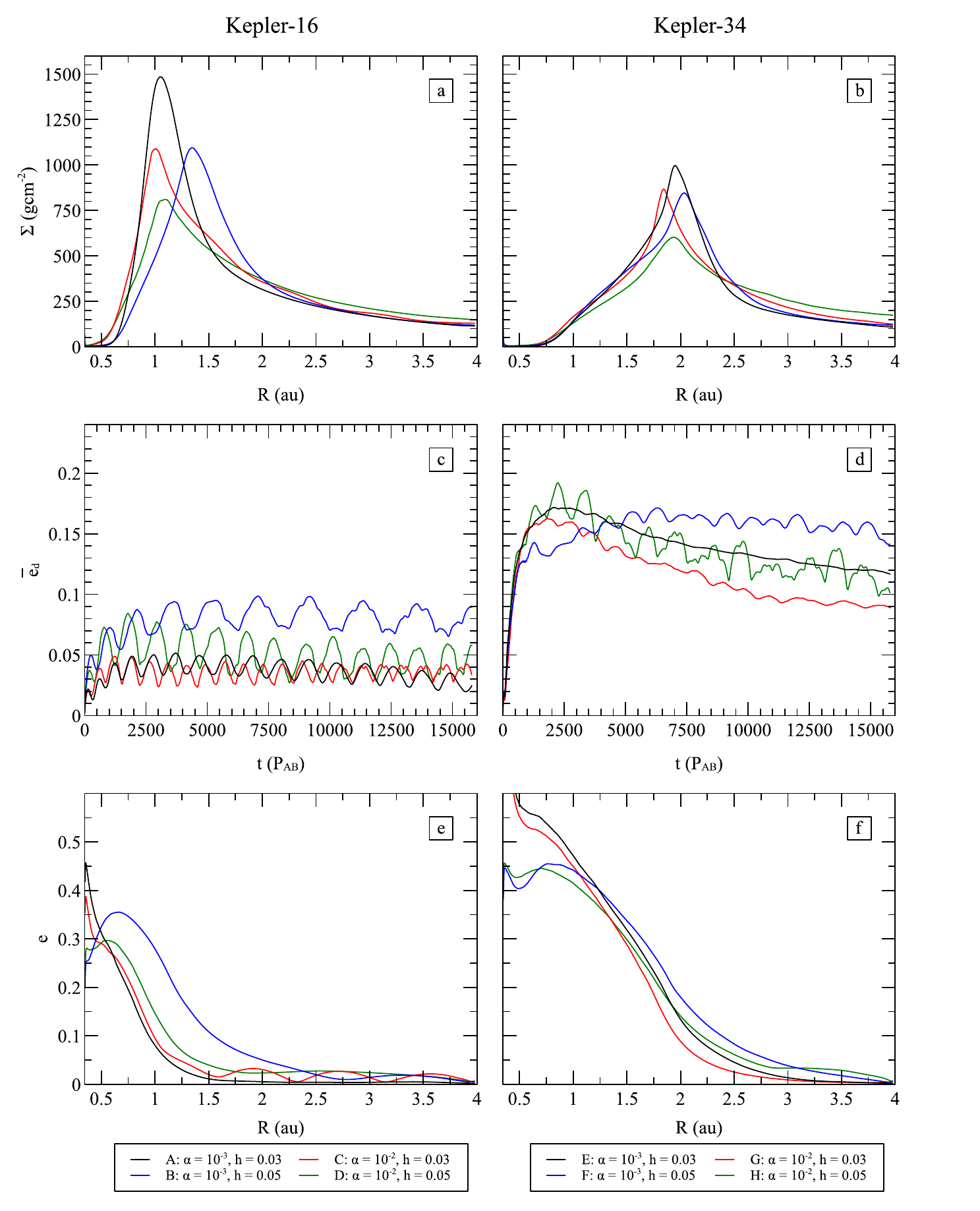}
\caption{Time-averaged quasi-steady state surface density profiles (a \& b), $\Sigma$, and time evolution of disk mean eccentricity (c \& d), $\bar{e}_d$, and time-averaged radial eccentricity profiles e \& f), e, for Kepler-16 (Models \emph{A - D}) and Kepler-34 (Models \emph{E - H}) under non self-gravitating conditions. Time-averaging is over the last 1,000 binary orbits of the simulations.}
\vspace{+5pt}
\label{fig:massive}
\end{figure*}

The initial surface density of the gas disk follows $\Sigma(R) = f_{gap}\Sigma_0R^{-\alpha_\Sigma}$ from PN13, where $\Sigma_0$ is the surface density at 1 au, $R$ is the radial position in au, and $\alpha_\Sigma$ defines the initial surface density gradient. For the majority of the simulations (Table \ref{tab:fargomodels} simulations $A$-$N$ \& $U$) we assume a half minimum-mass solar nebula, thus, $\Sigma_0 = 10^{-4}$ M$_\odot$/au$^2$ \citep{hayashi81}. In simulations $O$-$T$ $\Sigma_0$ is increased by a factor of 5, 10 and 25. The density gradient, $\alpha_\Sigma$ is set to 1.5 in all cases except the final simulation, $U$, where a shallower value of 0.5 is used for comparison to previous work \citep{marzari13}. In all simulations, the total central stellar mass is normalised to 1.0 M$_\odot$ but the surface density of the disk is not scaled to account for the physical mass difference between Kepler-34 and Kepler-16. Therefore, although the total physical mass of the disk is fixed, the disk mass relative to combined stellar mass changes.

To avoid numerical instabilities caused by extremely low density fluid cells in the inner cavity due to the torque of the binary on the disk inner edge, a density floor is added such that the minimum value, $\Sigma_{min}$, is set to 10$^{-9}$ of the initial density. This means that mass is not strictly conserved in regions of very low density.

The aspect ratio, $h = \frac{H}{R}$, where $H$ is the pressure scale height, is constant across the disk and varies between each model from 0.03 to 0.05. The aspect ratio does not determine the physical thickness of the disk, since the simulations are two dimensional, but defines the sound speed, $c_s = v_k h$, where $v_k$ is the Keplerian speed. Turbulence in the disk, a likely contribution of the MRI effect \citep{balbus91}, is explored by varying an $\alpha$-viscosity between 10$^{-3}$ and 10$^{-2}$.

Since our disks are low in mass, we expect them all to be linearly stable according to the Toomre stability criterion, $Q$ $>$ 1 \citep{toomre64}. For our half-MMSN density models, even near the inner edge of the disk where the density is at a maximum, the minimum Toomre value, $Q_{min}$, is 140. However, on consideration of the results of \cite{marzari09} that show self-gravity plays a significant role in circumprimary disks, self-gravity is included in simulations $I$-$L$ and $O$-$Q$.

Over 21 simulations, we explore the effect that changing the stellar binary parameters, aspect ratio, $\alpha$-viscosity, boundary condition and inclusion of disk self-gravity has on the evolution and quasi steady-state (QSS) profile of the disk. All simulations are evolved for 16,000 binary orbits, $P_{AB}$, equivalent to 1,700 orbits at 1 au, with the exception of $M$ $\&$ $N$ which run for 4,000 $P_{AB}$ to test the effect of the chosen inner boundary condition on the initial response of the disk.

\section{Results}\label{sec:results}

\subsection{Non self-gravitating disks}

\subsubsection{Kepler-16}

We first look at the response of a non self-gravitating disk to a stellar binary with parameters chosen to match the observed properties of the system Kepler-16(AB). The azimuthally and time averaged quasi-steady state surface density is shown in Figure \ref{fig:massive}a. The density maximum or peak, $\Sigma_{peak}$, indicates the location of the truncated inner edge, although the eccentric morphology of the disk requires a full 2-dimensional density map, $\Sigma_{2D}$ (Figure \ref{fig:panel}), to identify the edge precisely as a function of azimuth angle. Therefore, we will consider the location of $\Sigma_{peak}$ to be at the average truncation radius, $\bar{r}_t$. The value of $\bar{r}_t$ lies between 1.0 and 1.4 au which is consistent with what was found by PN13.

Increasing the aspect ratio from 0.03 to 0.05 raises the disk eccentricity (Figures \ref{fig:panel}c \& \ref{fig:panel}e), particularly for $\alpha = 10^{-3}$, which can also be seen in the surface density profiles by an increase in the value of $\bar{r}_t$. The more circular form of the disk in model $A$ leads to a higher averaged surface density, since more mass orbits at a similar orbital radius.

The eccentric nature of the disk is better shown by the time evolution of the mean disk eccentricity (Figure \ref{fig:massive}c), $\bar{e}_d$, which we define according to \cite{pierens07} as
\begin{equation}
\bar{e}_d=\frac{\int^{2\pi}_0 \int^{R_{out}}_{R_{in}}\Sigma_c e_c R \,\mathrm{d}R\,\mathrm{d}\phi}{\int^{2\pi}_0 \int^{R_{out}}_{R_{in}}\Sigma_c R \,\mathrm{d}R\,\mathrm{d}\phi},
\end{equation}
where $R_{in}$ and $R_{out}$ are the disk inner and outer radii and $\Sigma_c$ and $e_c$ are the fluid element surface density and eccentricity respectively. Note $e_c$ is calculated assuming the cell is orbiting the binary barycentre alone (a two body problem). Indeed, models with $h$ = 0.05 show a higher value of $\bar{e}_d$ at times both early in evolution and quasi-steady state (QSS), particularly for the low viscosity run. Our final disk eccentricity values lie between 0.03 and 0.08, the spread matching closely that found in PN13. However, we observe much larger values of eccentricity oscillation amplitude at QSS, which varies between models from $7.5 \times 10^{-3}$ to $1.3 \times 10^{-2}$ as opposed to a consistent value of $5.0 \times 10^{-3}$ found by PN13. This may be due to our smaller value of initial surface density. The increase in eccentricity with aspect ratio is also seen in the radial profiles of the disk's eccentricity (Fig.~\ref{fig:massive}e). The models with a larger aspect ratio (models $B$ and $D$) have a larger average eccentricity than those with a smaller aspect ratio (models $A$ and $C$) from the inner boundary until about 2.5 au.

\begin{figure}[t]
\centering
\vspace{+0pt}
\hspace{-15pt}
\includegraphics[scale=0.5]{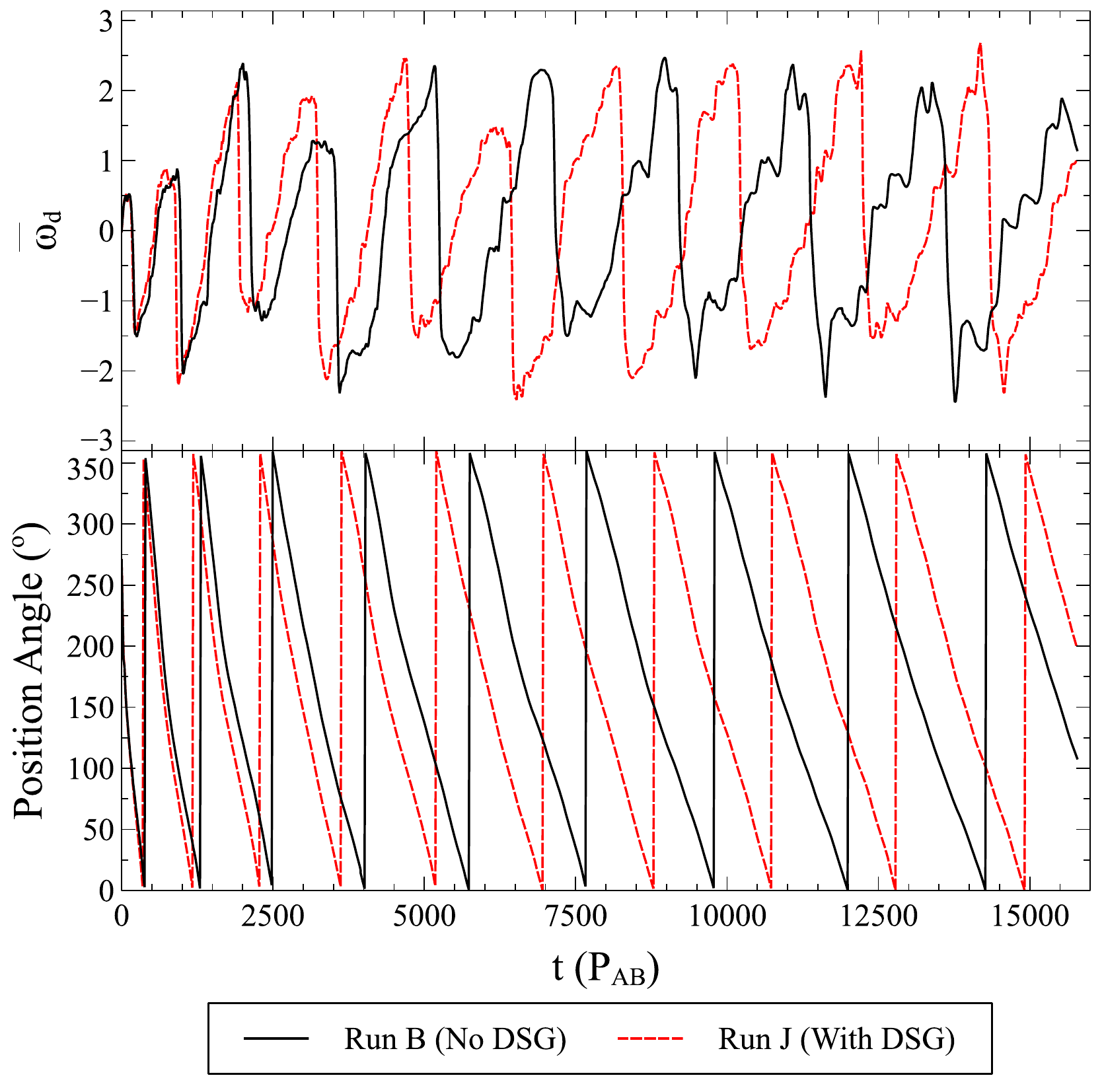}
\caption{Time evolution of disk mean longitude of periastron and the position angle of the disk centre-of-mass for Kepler-16 under both non self-gravitating conditions (black solid) and self-gravitating (red dashed) conditions.}
\vspace{+0pt}
\label{fig:B-pl}
\end{figure}

We can also calculate the disk mean longitude of periastron, $\bar{\omega}_d$, from the mass weighted average of the cell longitudes, $\bar{\omega}_c$:
\begin{equation}
\bar{\omega}_d=\frac{\int^{2\pi}_0 \int^{R_{out}}_{R_{in}}\Sigma_c \bar{\omega}_c R\,\mathrm{d}R\,\mathrm{d}\phi}{\int^{2\pi}_0 \int^{R_{out}}_{R_{in}}\Sigma_c R\,\mathrm{d}R\,\mathrm{d}\phi}.
\end{equation}
The value oscillates about the binary periastron which itself does not change due to the analytic prescription for the stellar orbits. In Figure \ref{fig:B-pl}, $\bar{\omega}_d$ is plot as a function of time for Kepler-16. On cross examination with the eccentricity time evolution (Fig.~\ref{fig:massive}c), it is clear that there is some relation between the eccentricity and periastron longitude since the maxima in the eccentricity oscillations matches the periastron longitude maxima. We discuss this link further in Section \ref{sec:dis}. The position angle of the disk centre-of-mass is calculated in the inertial frame and shows the circulation of the disk increasing in frequency until the disk precession period, $P_d$, settles at around 2,500 $P_{AB}$ at QSS.

The eccentric morphology of the disk is apparent through both $\Sigma_{2D}$ (Figure \ref{fig:panel}) and $\bar{e}_d$. Figure \ref{fig:panel2} shows the radial velocity which reveals a number of spiral features. In both the $h$ = 0.03 and 0.05 models of Kepler-16, the velocity map shows $m = 2$ spiral waves with a larger radial extent for the higher aspect ratio. To better assess the level of asymmetry and presence of modes in the disk, a Fourier analysis of the surface density is performed. 

The disk surface density can be decomposed into modes described by azimuthal mode numbers $m$ and frequency mode numbers $l$ since disk disturbances caused by the binary can have both an angular and time dependency. The surface density distribution can then be written as \citep{nixon15}

\begin{equation}
\Sigma(r,\theta,t)=\sum\limits_{l=-\infty}^{\infty}\sum\limits_{m=0}^{\infty}\mathrm{Re}[\Sigma_{l,m}(r)\exp[\mathrm{i}(m\theta-l\Omega_{b}t]],
\end{equation}
where $\Sigma_{l,m}(r)$ is a complex function and $\Omega_b$ is the mean motion of the binary, $2\pi/P_{AB}$. \cite{nixon15} find that eccentric binaries produce modes with $l \ne m$. In particular, in their retrograde circumbinary disk simulations, highly eccentric binaries ($e_b \ge 0.6$) produce powerful $l = -1$, $m = 2$ modes. We remove the time dependence of the disk disturbances by since our analysis is concerned with the eccentric Kepler-16 system ($e_b$ = 0.16) and because we focus only on a comparative study of the azimuthal mode strengths between simulations. We do this by choosing to discuss modes only with $l = 0$. This is done by performing the Fourier analysis of the surface density averaged over the period of a single binary orbit at QSS. This mode analysis allows for the identification of the mode propagation and dissipation properties with changing disk parameters.

\begin{figure}
\centering
\vspace{-0pt}
\includegraphics[scale=0.55]{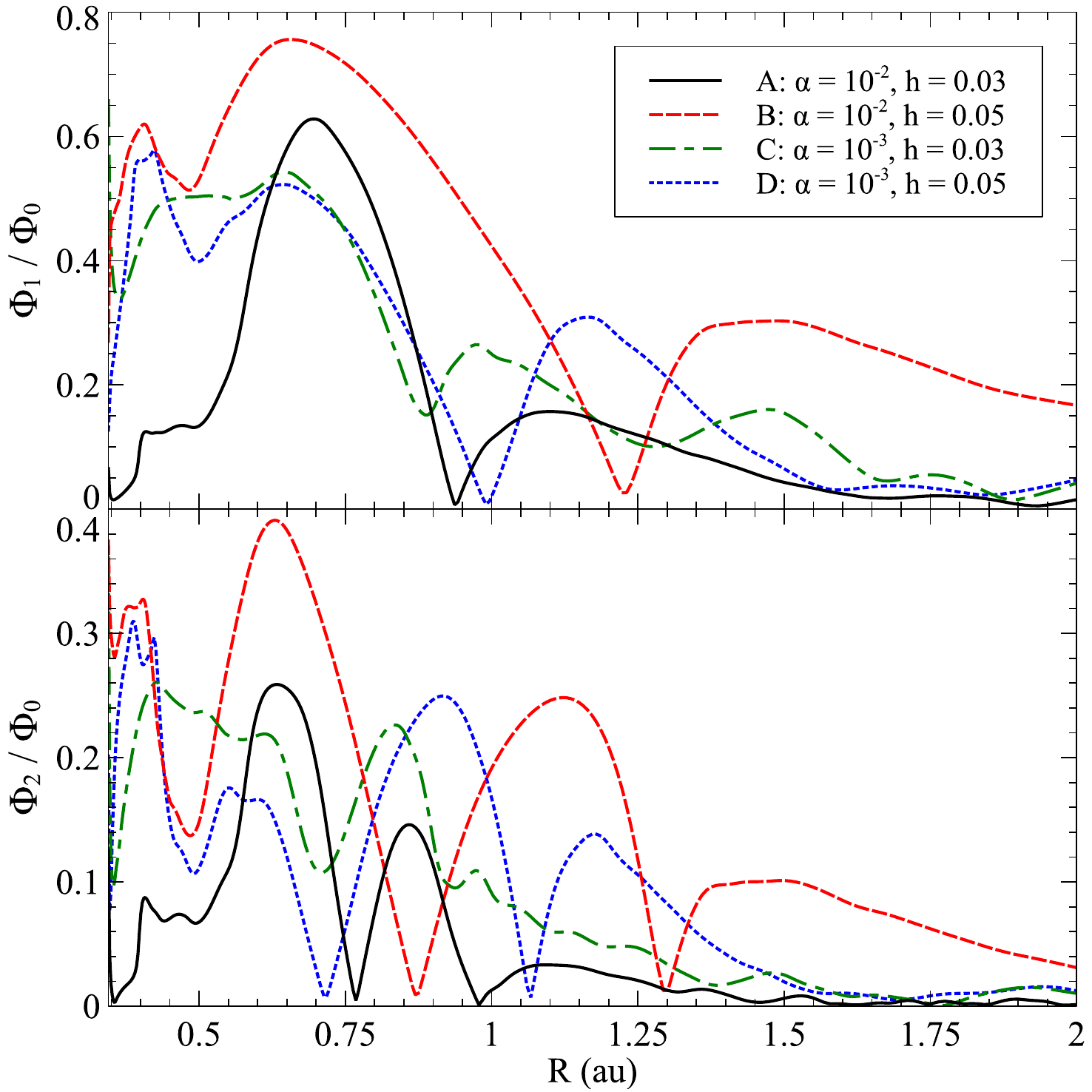}
\caption{Fourier analysis of the surface density of all non self-gravitating Kepler-16 circumbinary disks. The transform is done on the time-averaged surface density, over a single binary orbit at 15,000 $P_{AB}$, to consider only $l = 0$. The strengths of both the $m = 1$ and $m = 2$ modes are normalised against the axisymmetric ($m = 0$) contribution and plot as a function of radius in the disk.}
\vspace{+5pt}
\label{fig:hfourier}
\end{figure}

In Figure \ref{fig:hfourier} the strengths of the $m = 1$ and $m = 2$ modes relative to the axisymmetric component is plot as a function of orbital radius. The results confirm that for the high aspect ratio ($h$ = 0.05) models $B$ and $D$, the contribution from the $m = 2$ spirals is more significant in the outer disk than that seen for $h$ = 0.03.

\subsubsection{Kepler-34}

\begin{figure}[t]
\centering
\vspace{-0pt}
\hspace{-15pt}
\includegraphics[scale=0.6]{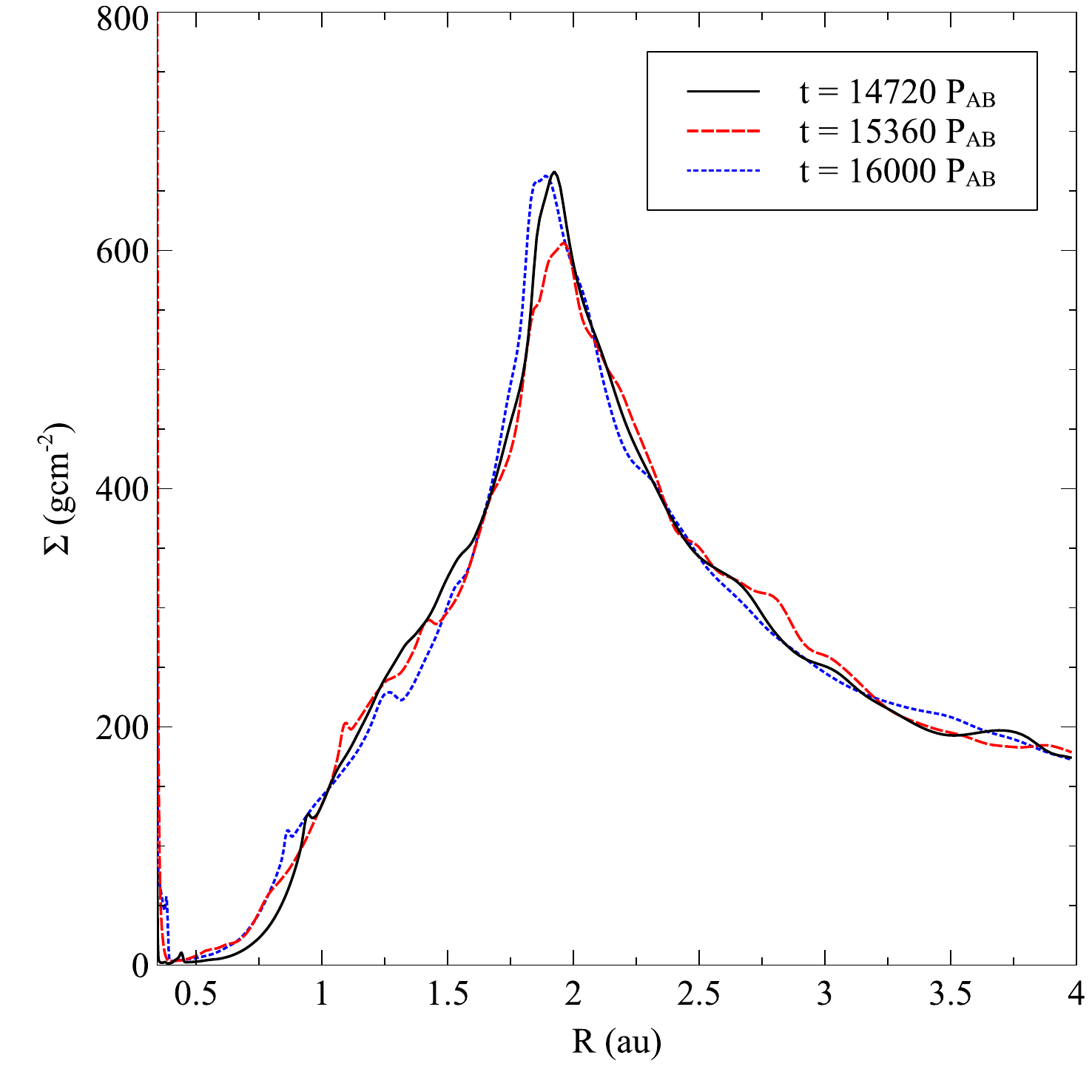}
\caption{Instantaneous surface density profiles for Model $H$ (Kepler-34) at three times near quasi-steady-state separated by 640 $P_{AB}$.}
\vspace{+5pt}
\label{fig:sgevo}
\end{figure}

Disks surrounding Kepler-34 are significantly more affected by the binary, with the QSS $\bar{e}_d$ ranging from 0.1 to 0.15, typically three times more eccentric than Kepler-16 (Figure \ref{fig:massive}d). Since the mass-weighted eccentricities decrease slightly with time by the simulation end, it is not clear from $\bar{e_d}$ if the disks have reached a steady state by 16,000 $P_{AB}$. However, looking at the instantaneous surface density profiles for a Kepler-34 run during the last few hundred binary orbits as shown in Figure \ref{fig:sgevo}, it appears as though a steady-state has been reached. Our Kepler-34 results agree roughly with PN13, who find that typically, for the systems mutually covered, a quasi steady-state is achieved by $10^4$ $P_{AB}$.

As with Kepler-16 the average eccentricity in the Kepler-34 disks once they have reached QSS is larger for higher values of aspect ratio. The large difference in $\bar{e}_d$ ($\Delta \bar{e}_d$ = 0.05) between models $F$ and $G$ is shown in Figure \ref{fig:massive}f to originate from the disk beyond the truncation radius where the surface density is much higher, and is therefore expected since $\overline{e}_d$ is a mass-weighted value. From 2.0 au, model $F$ has a small positive eccentricity offset until 3.0 au, but there is very little difference between all models interior to this location. There is a high level of agreement between aspect ratio and $\alpha$-viscosity models in the surface density profiles too; our simulations see $\bar{r_t}$ sharing a similar value of 2.0 au across all models. These results agree with PN13.

\subsection{Self-gravitating disks}

\begin{figure}[t]
\centering
\vspace{-0pt}
\hspace{-30pt}
\includegraphics[scale=0.65]{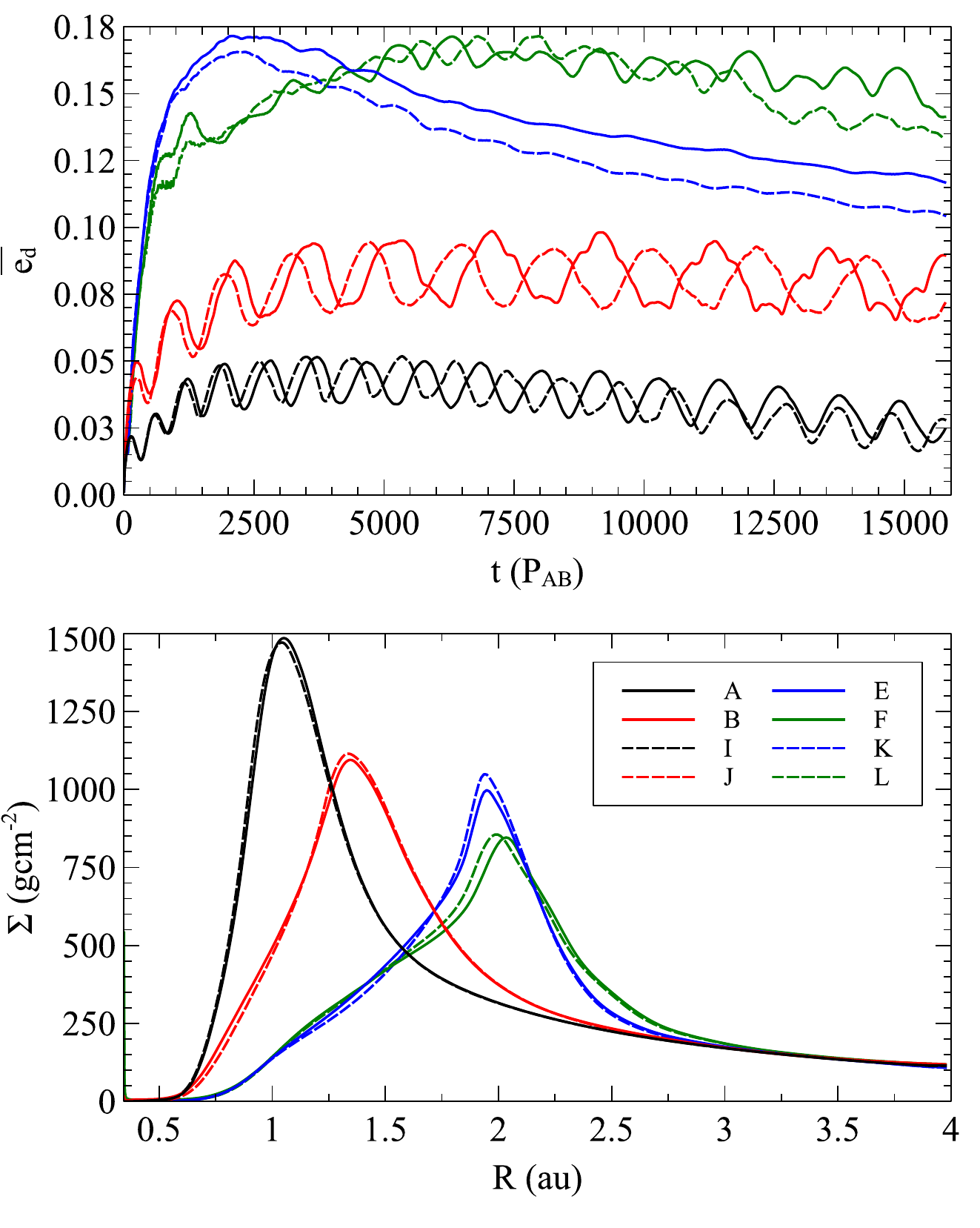}
\caption{Comparison of Kepler-16 disk eccentricity evolution and time averaged surface density for 0.5 $\Sigma_{MMSN}$ density disks with ($I$, $J$, $K$ \& $L$) and without self-gravity ($A$, $B$, $E$ \& $F$). Self-gravitating runs are shown with dashed lines.}
\vspace{+5pt}
\label{fig:sgmmsn}
\end{figure}

In models $I - L$ we enable disk self-gravity (DSG) for both Kepler-16 and Kepler-34. The $\alpha$-viscosity is fixed at $10^{-3}$ but the aspect ratio is varied between 0.03 and 0.05 to adjust the Toomre value and allow for direct comparison with models $A$, $B$, $E$ and $F$. For Kepler-16, as seen in Figure \ref{fig:sgmmsn}, during the first 2,000 binary orbits the disk responds almost identically to that of its non self-gravitating counterpart. The evolution up to QSS is subtly different however. The frequency of the eccentricity oscillations increases with gravity enabled suggesting that the disk circulation frequency increases. Due to the absence of strongly defined eccentricity oscillations in the Kepler-34 simulations, it is possible only to differentiate between self- and non self-gravitating models by the eccentricity magnitude, and not the oscillations. We find a slightly reduced disk eccentricity with DSG enabled for the larger aspect ratio simulation. In Figure \ref{fig:sgmmsn} the time averaged surface density profiles also reveal that a self-gravitating disk in Kepler-16 has no effect on the final disk structure, and almost no effect for disks around Kepler-34. 

To test at what disk mass self-gravity become important, $\Sigma_0$ is increased by 5, 10 and 25 times. Self-gravity importance is therefore tested for $Q$ $\rightarrow$ 1 but $Q$ $>$ 1 at all times. We define standard density models ($\Sigma_0$ = 1 $\times$ 10$^{-4}$ code units) as being 0.5 $\Sigma_{MMSN}$ and those scaled up by a factor of 5, 10 and 25 as 2.5 $\Sigma_{MMSN}$ ($Q_{min}$ = 28), 5 $\Sigma_{MMSN}$ ($Q_{min}$ = 14) and 12.5 $\Sigma_{MMSN}$ ($Q_{min}$ = 5) respectively.

\begin{figure}[t]
\centering
\vspace{+0pt}
\hspace{-15pt}
\includegraphics[scale=0.65]{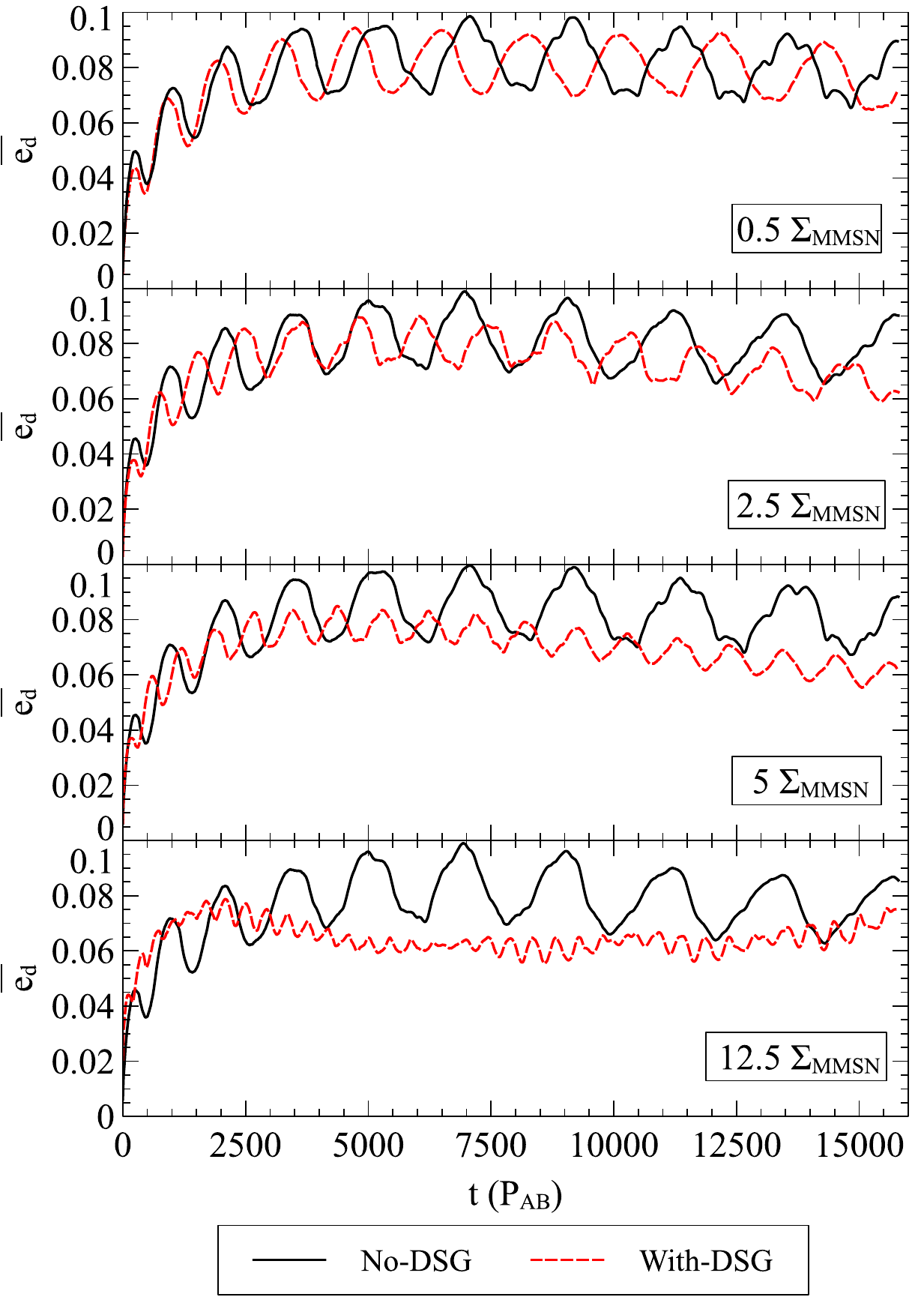}
\caption{Evolution of the mean disk eccentricity for standard density self-gravitating Kepler-16 run $J$ and increased density runs $O$-$T$ (dashed lines). The corresponding simulation with self-gravity disabled is shown by solid lines.}
\vspace{+5pt}
\label{fig:sget}
\end{figure}

In Figure \ref{fig:sget} the evolution of the mean disk eccentricity for each of the four varying densities is shown. $\bar{e}_d$ remains, for the non self-gravitating runs, at a constant 0.08 for each surface density. When self-gravity is enabled, with the exception of 12.5 $\Sigma_{MMSN}$, $\bar{e}_d$ falls off slowly with time and the rate of this fall off increases with increasing surface density. This leads to disks with higher surface densities having a slightly (5-25\%) smaller eccentricity at the end of the simulation. The evolution of the eccentricity in the self-gravitating 12.5 $\Sigma_{MMSN}$ simulation is more dynamic due to an eccentricity fall off before increasing again.

At standard densities we find that the oscillation frequency increases and amplitude decreases with self-gravity enabled. Increasing the density does not change the oscillation frequency without self-gravity enabled, but the period of the oscillations decreases from 2000 $P_{AB}$ (0.5 $\Sigma_{MMSN}$) to 300 $P_{AB}$ (12.5 $\Sigma_{MMSN}$) when it is included.

Another measure of the effect of self-gravity is to identify the strength of the modes present in the disk. A fourier analysis of the surface density, identical to that done for the non self-gravitating Kepler-16 runs, is done. The Fourier coefficients are determined and normalised against the axisymmetric ($m = 0$) component. The strength of these modes normalised to the axisymmetric part, $\Phi_m/\Phi_0$, is shown for each surface density in Figure \ref{fig:fc}. Again, a comparison is made between simulations with and without self-gravity included.

There is a strong contribution from both the $m = 1$ and $m = 2$ modes, with the $m = 1$ mode always 20$\%$ or larger in the inner disk. It is less clear as to how the inclusion of self-gravity changes the strength of these modes. When self-gravity is enabled in the 0.5 $\Sigma_{MMSN}$, 2.5 $\Sigma_{MMSN}$ and 5 $\Sigma_{MMSN}$ surface density runs, there are only subtle differences between the strength of the modes. At 12.5 $\Sigma_{MMSN}$, due to oscillations in the strength of both modes in the self-gravitating disk, the profiles for both the self-gravitating $m = 1$ and $m = 2$ modes do not trace their non self-gravitating equivalents.

\subsection{Surface density profile gradient}

\begin{figure}[t]
\centering
\vspace{-0pt}
\hspace{-30pt}
\includegraphics[scale=0.7]{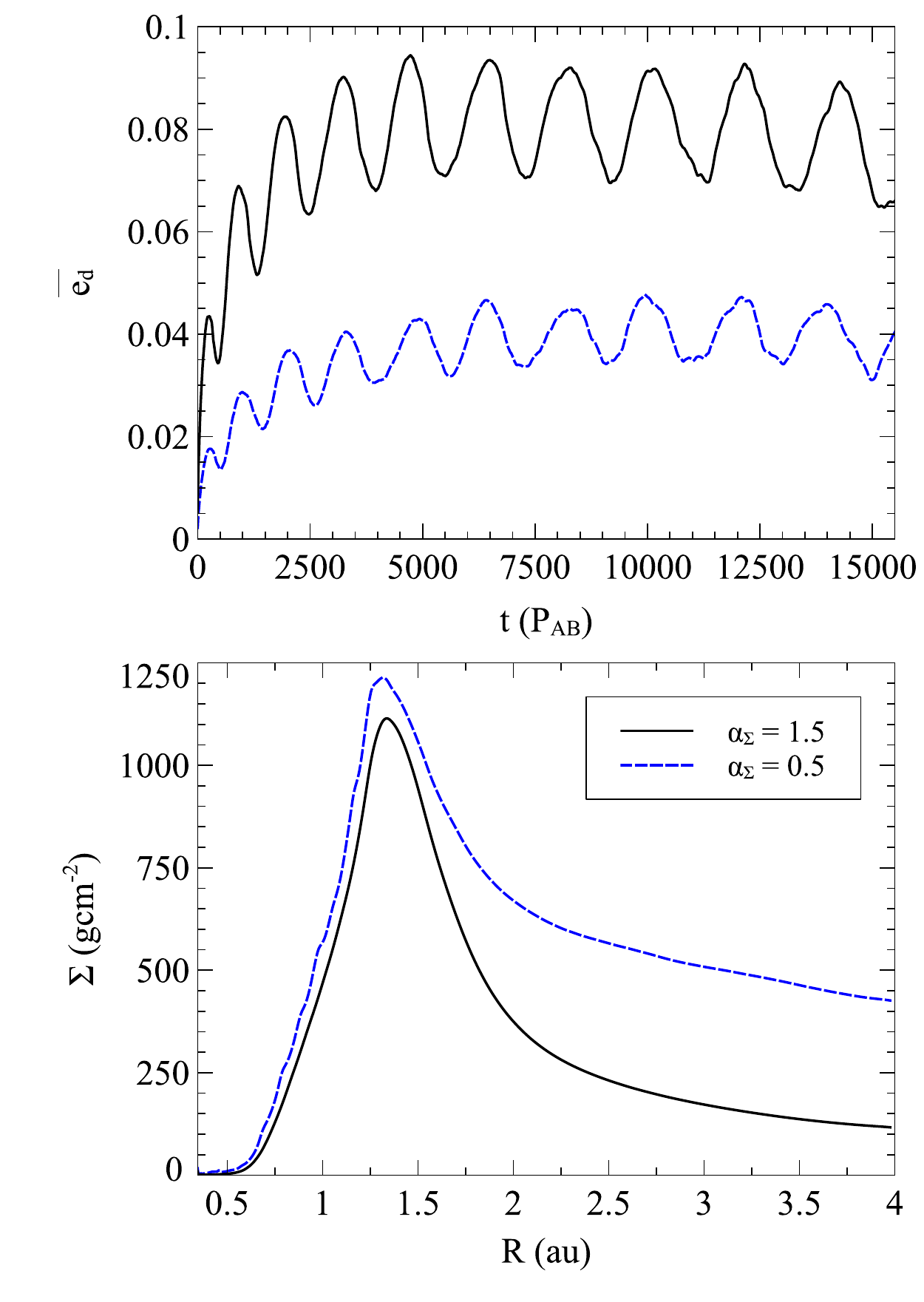}
\caption{Comparison of surface density and eccentricity evolution in Kepler-16 runs between density profiles of form $\Sigma$(r) $\propto$ $\Sigma_0$$r^{-\alpha_\Sigma}$ with $\alpha_\Sigma$ = 0.5 (blue dashed) and 1.5 (black solid)}
\vspace{+5pt}
\label{fig:dencomp}
\end{figure}

The gradient of the initial density profile is a value which often varies between studies, justified by the lack of observational data. In simulation $U$ the density gradient is made shallower by changing the surface density exponent $\alpha_\Sigma$ from 1.5 to 0.5. In Figure \ref{fig:dencomp}, the radially varying surface density is shown alongside the time evolution of the mean disk eccentricity. Changing the exponent has little effect on the truncation of the inner disk edge since $\Sigma_{peak}$ adopts the same value. However, there is a significant difference in the mean disk eccentricity at all times, with $\alpha_\Sigma$ = 0.5 obtaining a much lower eccentricity throughout.

\subsection{Boundary conditions}

Previous work on circumbinary disks sees the implementation of two boundary conditions for the inner edge; namely rigid/reflecting and open. Open boundary conditions are chosen to be the more realistic scenario since matter can flow through the boundary and accrete onto the star(s) as would occur in a real system. The difficulty in these cases however is setting a well defined value for $v_r$ and $v_\phi$. Rigid boundary conditions are chosen to remove the enhanced mass loss experienced through the inner disk from an eccentric disk. The rigid approximation is, however, unrealistic and is often modified to include routines that remove reflected waves at the boundaries. We explore the influence of the choice of boundary condition on the disk by comparing the surface density and eccentricity between identical simulations, but with different inner boundary conditions.

\begin{figure*}
\centering
\includegraphics[scale=0.9]{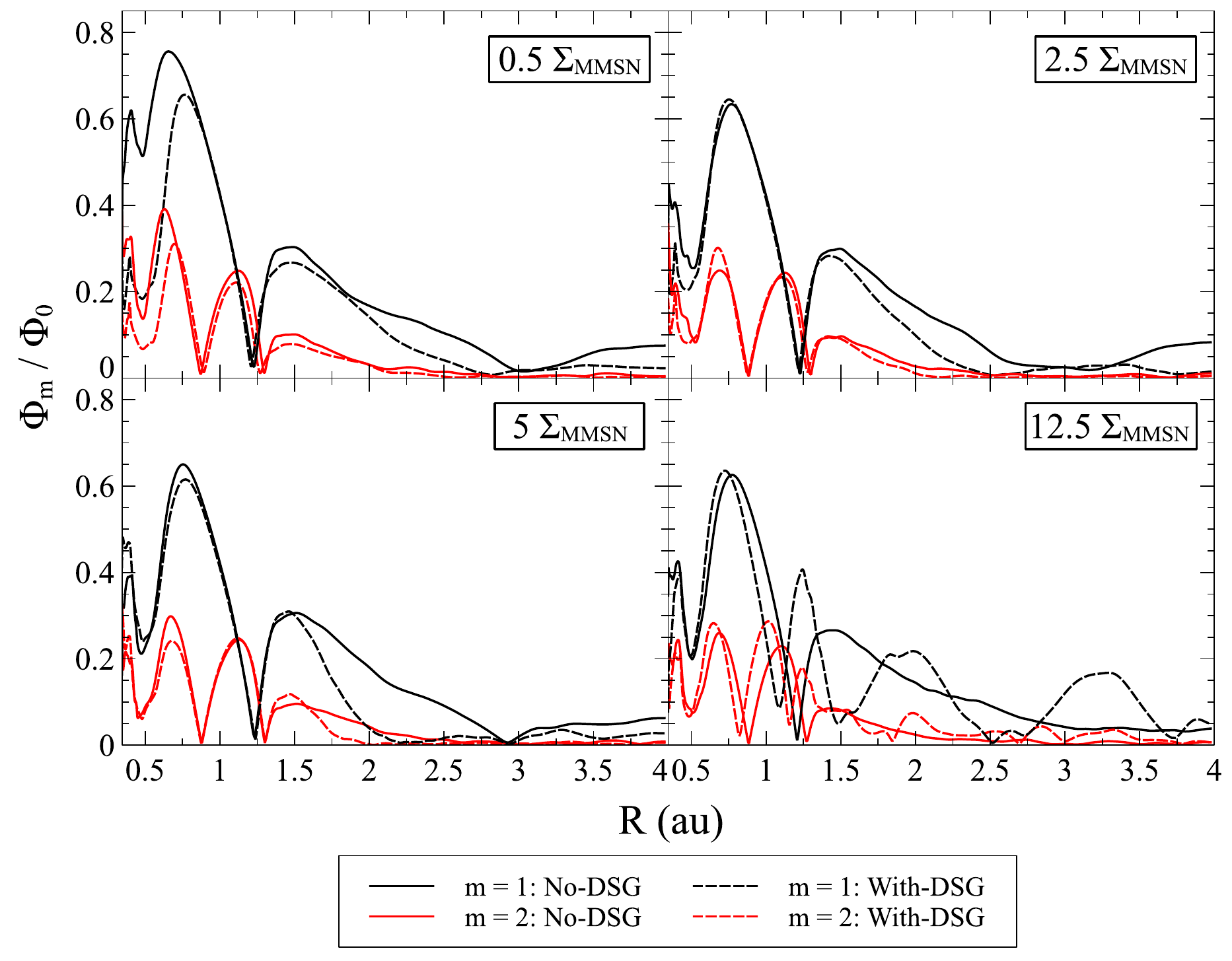}
\vspace{+10pt}
\caption{Fourier analysis of the surface density of Kepler-16 simulations testing the relevance of self-gravity with increasing $\Sigma_0$. Black lines correspond to the Fourier coefficients of the $m = 1$ modes, normalised to the axisymmetric $m = 0$ component. Red lines show the $m = 2$ modes. Simulations where self-gravity is enabled are shown by dotted lines. Fourier analysis is done on the time-averaged surface density over the period of one binary orbit from 15,000 $P_{AB}$.}
\vspace{+10pt}
\label{fig:fc}
\end{figure*}

\begin{figure}
\centering
\vspace{-0pt}
\hspace{-25pt}
\includegraphics[scale=0.65]{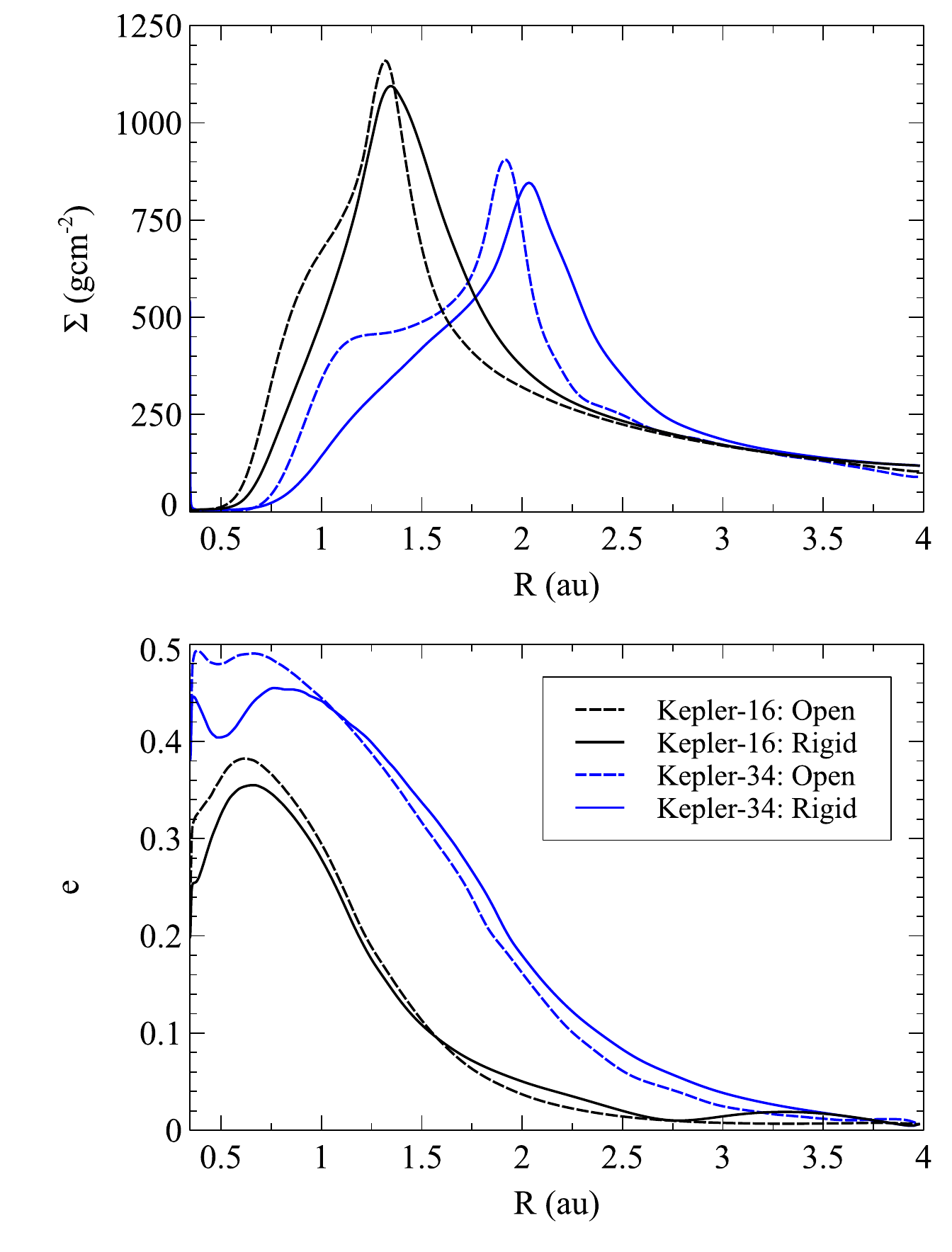}
\caption{Comparison of open (dashed) and closed (solid) boundary conditions in Kepler-16 (black) and Kepler-34 (blue) via surface density and eccentricity profiles.}
\vspace{+5pt}
\label{fig:boundary}
\end{figure}

We find that there are small differences in both the disk eccentricity and structure between the two. In Kepler-16 the surface density profiles show that including an open boundary keeps the peak density at the same location but increases the mass interior to $\Sigma_{peak}$. This effect, which is shown in Figure \ref{fig:boundary} is stronger in Kepler-34 where the density plateaus at around 1.1 au to 1.5 au before further increasing to a peak at 2 au. We present our explanation for this in the Section \ref{sec:dis}. The disk eccentricity reveals less of a difference between the two boundary conditions. In the main part of the disk the eccentricity is the same regardless of boundary condition although the mass weighted eccentricity is higher in the cavity.

\section{Discussion}\label{sec:dis}

In this paper, we have shown the results of a series of hydrodynamical simulations that explore the response of a circumbinary gas disk to both the Kepler-16 and Kepler-34 stellar binary systems. We vary  $\alpha$-viscosity and aspect ratio, as well as probing the effects of disk self-gravity, initial surface density profile, and the inner boundary condition.

The majority of our simulations have reached steady state (a consistent value of $\bar{e}_d$) by 16,000 $P_{AB}$. The steady-state nature of the disk is also reflected in the stationary form of the instantaneous, azimuthally averaged surface density during the last few thousand binary orbits (Figure \ref{fig:sgevo}), showing the balance between the truncation effects from tidal torques on the inner edge of the disk and the viscous replenishment of mass to the inner disk. The formation of the central cavity, which is shown by the movement of $\Sigma_{peak}$ in Figure \ref{fig:panel}, causes the inner edge of the disk to move outwards. By taking the cavity edge to be the location at which the positive density gradient begins, we show good agreement with the \cite{artymowicz94} gap estimate $R_{gap} = 2.5a_b$.

\begin{figure}[t]
\centering
\vspace{-0pt}
\hspace{-15pt}
\includegraphics[scale=0.6]{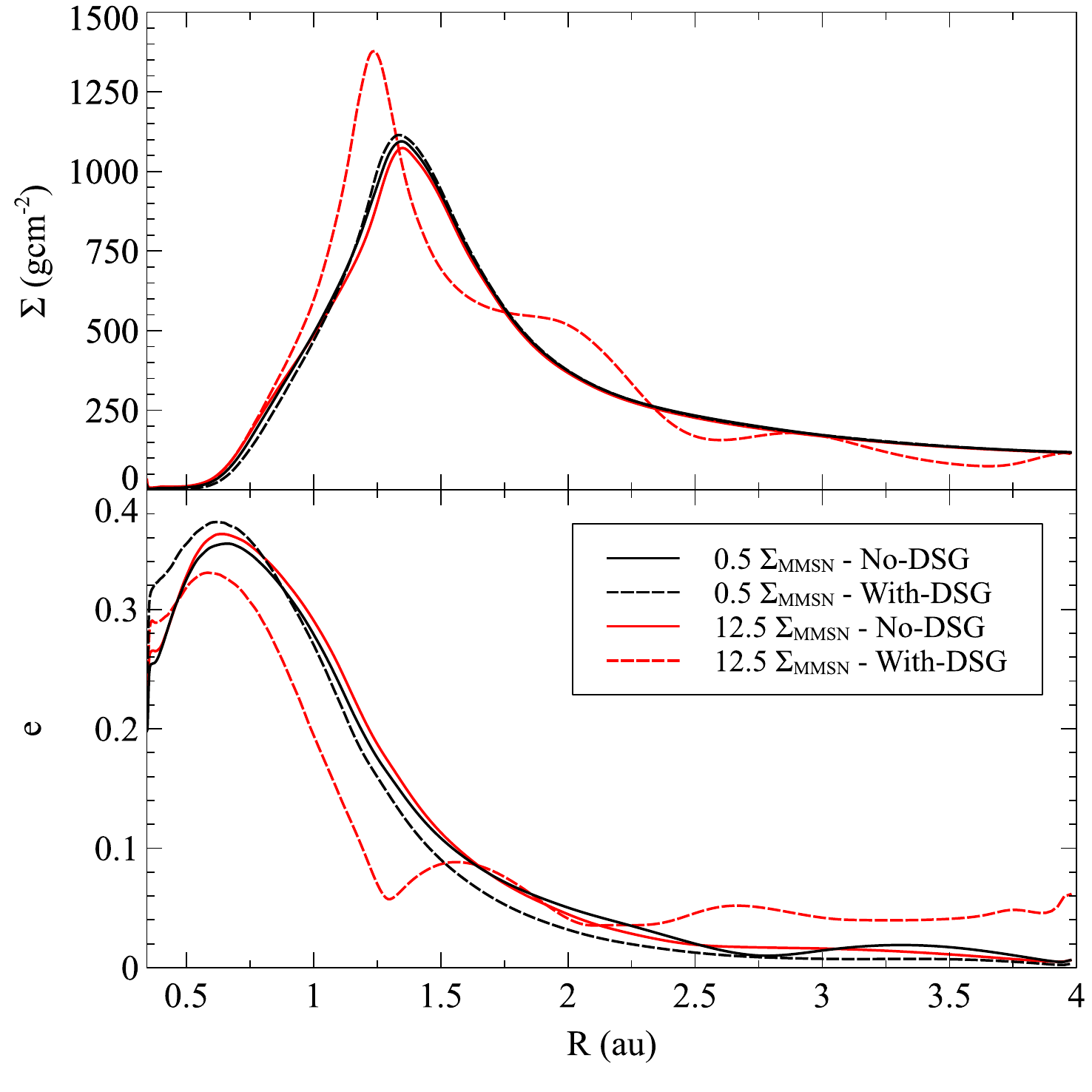}
\caption{Comparison of surface density and eccentricity profiles in Kepler-16 runs between self-gravitating (dashed) and non self-gravitating (solid) for 0.5 $\Sigma_{MMSN}$ (black) and 12.5 $\Sigma_{MMSN}$ (red). The 12.5 $\Sigma_{MMSN}$ surface density profile is normalised to the initial half minimum-mass solar nebula surface density.}
\vspace{+5pt}
\label{fig:sgcomp}
\end{figure}

The results for the surface densities and eccentricities of our non self-gravitating disks agree with those found by PN13. Subtle differences such as the eccentricity magnitude (Figure \ref{fig:massive}), particularly in Kepler-34, may be explained by a number of factors that include our reduced outer boundary (from 5 au to 4 au) and the difference in initial surface density profiles. We certainly retrieve higher disk eccentricities than found by \cite{kley14} who attribute their lower values in comparison with PN13 to the choice of boundary condition. \citet{kley14} claim that the rigid boundary condition used by PN13 leads to a higher disk eccentricity. We find that the choice of a rigid boundary condition leads to only the slightest increase in disk eccentricity over the open boundary scenario in Kepler-34, and leads to lower disk eccentricities for $a$ < 1.5 au in Kepler-16. Therefore, we suggest that it might be more appropriate to link the discrepancy in disk eccentricity to the choice of stellar binary for which we have shown significantly affects the disk structure and dynamics. 

The shape and strength of the binaries potential invariably changes with its orbital properties: mass ratio, eccentricity and semi-major axis. Therefore, it is likely that a circumbinary gas disk is only weakly affected by the relatively low eccentricity of Kepler-38 (e = 0.10) \citep{orosz12a} compared to the significantly higher eccentricity of Kepler-34 (e = 0.53). We would therefore expect the gas disk around Kepler-34 to adopt a higher eccentricity than a similar disk around Kepler-38. This theory is supported by the significantly larger disk eccentricities in simulations around the eccentric Kepler-34 compared to Kepler-16 where the stellar binary has a much smaller eccentricity.

Increasing disk eccentricity with increasing stellar binary eccentricity is a trend also seen in $N$-body results. The planetesimal response in the \emph{N}-body scenario is dependent upon the $forcing$ of stellar binary on the disk, and perturbation theory provides a value of eccentricity of which planetesimals tend towards. This forced eccentricity, $e_f$, is given by \citep{moriwaki04}
\begin{equation}
e_f = \frac{5}{4}\frac{M_A - M_B}{M_*}\frac{a_b}{R}e_b\frac{1+3e_b^2/4}{1+3e_b^2/2},
\label{eq:ef}
\end{equation}
where $M_A$ and $M_B$ are the primary and secondary stellar masses, $M_*$ is the total binary mass, $R$ is the distance from the binary barycentre and $a_b$ and $e_b$ are the binary semi-major axis and eccentricity, respectively. While it is clear that $e_f$ varies with $e_b$ from Equation \ref{eq:ef}, how $e_d$ is precisely affected by $e_b$ is not known since Kepler-34 and Kepler-16 differ in mass ratio as well as stellar binary eccentricity. The remaining question is do the gas and planetesimal disks adopt the same eccentricity; are $e_f$ and $e_d$ equal? On average we find at 1 au, $e_d \sim 0.2$ for Kepler-16 (as per Figure \ref{fig:massive}e), whereas $e_f \sim 0.02$. Kepler-34 has an even larger difference between its value of $e_d \sim 0.45$ and $e_f \sim 0.002$. 

There is a noticeable trend in $\bar{e}_d$ with aspect ratio but no clear change in the disk eccentricity or structure with varying $\alpha$-viscosity. We find considerably higher eccentricities in disks with a higher aspect ratio. This is explained by the increased level of communication in the disk fluid due to the larger value of the sound speed, $c_s$. The higher $c_s$ allows waves in the disk to propagate over longer radial distances before being damped. This results in an increased pitch angle of the spiral arms and increases the radial extent to which the waves can travel. Waves which unfold further out can deposit their energy over a larger percentage of the disk which increases the mass-weighted mean disk eccentricity, $\bar{e}_d$ (see Fig.~\ref{fig:panel2}). This effect is seen in Figure \ref{fig:hfourier} as the enhanced propagation from a larger aspect ratio leads to a more significant contribution of $m = 2$ modes in the outer regions of the disk. The deposition of energy as a result of these unfurling spirals in the large aspect ratio runs, is also indirectly observed through the enhanced eccentric $m = 1$ modes in the outer disk, as the binary transfers energy to the fluid elements. Additionally, the radial eccentricity profiles of Kepler-16 show that a high aspect ratio leads to eccentricity increases in the outer regions of the disk. This can be seen in Figure \ref{fig:massive}e, where the eccentricity beyond 1.5 au in Kepler-16 is close to 0 for the low aspect ratio runs $A$ and $C$, but remains greater than 0 for both runs $B$ and $D$. This is also the case for Kepler-34 which has e = 0.05 at 3 au for both high aspect ratio runs $F$ and $H$, but close to 0 for the low aspect ratio runs $E$ and $G$.

There is a correlation between oscillations in the disk eccentricity and the rotational dynamics of the disk from cross examining the longitude of periastron in Figure \ref{fig:B-pl} and the disk eccentricity in Figure \ref{fig:massive}. \cite{pierens13} suggest that the amplitude of the oscillations is linked to pressure effects. This is reflected in our results by the increasing amplitude of eccentricity oscillations with larger aspect ratio. For a disk around an eccentric binary \cite{lubow00} find the direct forcing from the $m = 1$ binary potential ($\Phi_{1,1}$) on the disk can become important and the growth of the disk eccentricity is described (their Equation 2) as

\begin{equation}
\dot{e}_d = -\frac{15}{16}e_b\mu_b(1-\mu_b)(1-2\mu_b)(a_b/a_d)^{3}(1-e_d^2)^{-2}\Omega_dsin{\varpi},
\label{eq:egrowth}
\end{equation}
where $e_b$, $\mu_b$ and $a_b$ are the binary eccentricity, mass parameter ($M_B/M_t$) and semi-major axis respectively, $\Omega_d$ is the Keplerian velocity of the disk and $\varpi$ is the longitude of periastron of the disk relative to the binary. We do not retrieve the same magnitude of eccentricity growth as predicted by Equation \ref{eq:egrowth} but this is likely due to the difference between the model's dependency on ballistic particles in contrast to our fluid. The eccentricity growth and damping is well phased  though with $\varpi$, mapping the eccentricity oscillations in Kepler-16. Since the binary is placed on a fixed orbit $\bar{\omega_d} \equiv \varpi$. Therefore by reading the evolution of the longitude of periastron of run B in Figure \ref{fig:B-pl}, during periods where $\varpi < 0$, Equation \ref{eq:egrowth} gives $\dot{e}_d > 0$ and hence eccentricity growth. These periods are matched by an eccentricity increase for run B in Figure \ref{fig:massive}c. Similarly regions where $\varpi > 0$ and hence $\dot{e}_d < 0$, a damping of eccentricity is seen in the eccentricity plot. The comparison of the phasing of $\dot{e}_d$ with $\varpi$ for Kepler-34 is difficult since the eccentricity oscillations in disks around this binary are less pronounced. This validates the model further however, as $\Phi_{1,1} = 0$ for Kepler-34, since the stars have equal mass ($\mu_b =  0.5$). A similar effect is observed in the $N$-body case where the mass ratio of Kepler-34 leads to the removal of secular forcing ($e_f = 0$ and dynamical forcing $e_{ff} \neq 0$) \citep{paardekooper12}.

There is a discrepancy between the centre-of-mass position angle showing perfect circulation of the disk and the longitude of periastron showing a combination of imperfect circulation and libration. This may be explained by the mass weighted value of the longitude of periastron becoming contaminated by outer regions of the disk where higher order disturbances can cause the value of $\bar{\omega}$ to flip such that the orientation of the eccentric disk can vary between the inner and outer regions. If $\bar{\omega}$ remained on the same hemisphere of the disk for all radii then $\bar{\omega}_d$ would extend fully from -$\pi$ to +$\pi$ to show full circulation as per the centre-of-mass.

We find a very minimal contribution from self-gravity to the dynamics of the disks. At standard densities, in both Kepler-16 and Kepler-34 the eccentricity is marginally lower with gravity enabled, an effect observed by \cite{marzari09} in their study of circumstellar disks perturbed by an s-type binary (circumprimary) configuration. Additionally, the oscillation frequency in $\bar{e}_d$ increases with gravity enabled due to the increase in precession rate of the disk as seen in Figure \ref{fig:B-pl}. Both these effects have only a minor impact on the disk's density profile and its eccentricity. Therefore including disk self-gravity in simulations of disks at half minimum-mass solar nebula size is not necessary in modelling the disk and retrieving the correct eccentricity and structure. 

We also find that self-gravity has no real influence on the evolution of the magnitude of the disk eccentricity or steady-state surface density at 2.5 and 5 times the minimum-mass solar nebula. At 12.5 times, self-gravity starts to become a necessary addition to the simulations to resolve the correct morphology of the disk. Enabling it at this high disk mass changes the structure of the disk, as can be seen in the surface density plot of Figure \ref{fig:sgcomp}. A series of small density humps in the outer disk is observed that occur at regions of increased disk eccentricity. A noticeable difference in the disk evolution when self-gravity is included is the change in frequency of the eccentricity oscillations. The trend in final mass weighted mean disk eccentricity is to decrease with increasing surface density, with the exception of the highest density run. The disk eccentricity is probed further in Figure \ref{fig:sgcomp} where the eccentricity of the inner disk is greatly reduced with self-gravity enabled for 12.5 $\Sigma_{MMSN}$. The eccentricity then increases over the non self-gravitating run for $R$ $>$ 2.0 au. This may explain why the mass weighted value is larger at the end of the simulation, since it takes time to raise the eccentricity of the outer disk. Including self-gravity therefore acts to slightly damp the overall eccentricity of the disk. Global self-gravitating modes can exist in the disk if precession due to the companion star can be ignored \citep{papaloizou02}. However, it is likely that disk precession due to a massive binary companion cannot be ignored, in which case the global modes play no role of importance.

The Fourier analysis performed to identify differences in the strength of the low frequency modes did not reveal any significant trends with increasing surface density, with or without self-gravity included (Figure \ref{fig:fc}). With the exception of 12.5 $\Sigma_{MMSN}$ where including self-gravity is required to accurately model the $m = 1$ modes in the outer disk, it does not appear that including self-gravity is important when considering disks with $Q$ $>$ 14. Regardless of the surface density magnitude, the results reveal the importance of including up to the $m = 2$ contribution of the surface density in fully describing the structure of the disk since outside the cavity, within the disk, the strength of the $m = 2$ modes is often 20\% of the background axisymmetric surface density. 

Our results indicate that the choice of boundary condition does not reflect strongly in the dynamical evolution of disks around both binary types. In both binary configurations, allowing material to flow through an open boundary increases the mass interior to the density peak. This effect is stronger in Kepler-34 than Kepler-16. Explanations for this include advanced wave damping routines operating in the rigid boundary case that could change the structure of the inner disk, or that the open boundary allows for a more realistic truncation timescale. There is little difference in the disk eccentricities beyond the truncation radius. These results are in contrast with \cite{kley14} who find much larger disk eccentricities when using a rigid boundary. This is probably to be expected since their shallower initial surface density profile ($\Sigma \propto r^{-1/2}$) and omission of a gap function to model the expected inner cavity means that a large amount of disk mass is located just exterior to the inner boundary. This means a large disk mass is perturbed close to the binary, increasing disk eccentricity. In this case, an open boundary is quick to remove any highly eccentric material close to the boundary itself.

The final result from this work is the effect the initial surface density gradient, $\alpha_{\Sigma}$, has on the eccentricity. By changing $\alpha_{\Sigma}$ from 1.5 to a shallower gradient of 0.5, we retrieve a much smaller eccentricity at all times. Using $\alpha_{\Sigma}$ = 0.5 reduces the final disk eccentricity by a half. This result is not surprising since the calculation of the mean disk eccentricity is mass weighted and therefore biased towards areas of high mass. For $\alpha_{\Sigma}$ = 0.5, more mass is in the outer disk, as seen in the surface density profiles of Figure \ref{fig:dencomp}. Since the outer disk is less perturbed by the binary and has a lower eccentricity, disks with a shallower density profile will have a lower mass weighted eccentricity. This explains why, alongside the varying binary parameters, \cite{kley14} find considerably smaller values for their disk eccentricities than found by us and PN13.

\section{Summary and further work}\label{sec:summary}

In this paper we have performed a number of hydrodynamical simulations of circumbinary gas disks that explore the binary configuration, disk aspect ratio, $\alpha$-viscosity, inner boundary condition, initial surface density profile and inclusion of self-gravity on the evolution and quasi steady-state of such disks.

We found that the choice of stellar binary has a significant impact on the eccentricity of the disk with Kepler-34 pumping up the mean disk eccentricity to three times the value seen in some simulations of Kepler-16. The truncation radius is also strongly affected by the binary configuration, with the more eccentric Kepler-34 forcing the inner edge out to an average radius of 2.0 au, compared to 1.0 au in Kepler-16. There is no significant correlation between the $\alpha$-viscosity and the final surface density but there is a noticeable trend for the disk eccentricity to increase with increasing aspect-ratio. The eccentricity of the disk is strongly linked to the initial surface density gradient, which must therefore be taken into account when comparing work of a similar nature. At least for our initial surface density profile gradient value of $\alpha_{\Sigma}$ = 1.5, the choice of inner boundary condition did not have a substantial effect on the evolution of the disk for both Kepler-34, and Kepler-16, but may play a role in resolving the disk interior to the density peak around the truncation radius. 

We did not see a strong influence of self-gravity in disks near minimum-mass solar nebula size, and saw only a minimal decrease in mean disk eccentricity when self-gravity is included. Fourier analysis of the surface density to reveal the presence of modes also suggested that enabling self-gravity, even in disks where the surface density causes the Toomre parameter $Q$ to approach 1, does not change the structure of the disks. We saw a subtle difference however in the $m = 1$ mode strength and variation in our high mass (12.5 $\Sigma_{MMSN}$) model when self-gravity is included, suggesting that work on disk masses larger than this value should include self-gravitation in order to accurately resolve the structure of the disk. This contrasts results showing the relevance of self-gravity in the circumprimary case since we did not find that it is a necessary inclusion in circumbinary disks that lie above the Toomre stability limit.

The next goal is to take surface density and velocity profiles of a steady-state disk for both Kepler-16 and Kepler-34 for use in gas potential feedback on our \emph{N}-body simulations of planetesimal evolution and growth. From this work we found that many of our parameter choices have little influence on the final product, with the exception of aspect-ratio and initial surface density gradient which both have a significant impact on disk eccentricity. This will be an important aspect to consider in our future work since the eccentricity of the gas disk is a measure of the level of asymmetry in the gas disk potential.

Future work on the hydrodynamical evolution of circumbinary disks will include the thermal response of the gas by including an energy equation. Additionally, since the evolution of the disks is sensitive to binary configuration, further investigations should look at changing the binary parameter space: eccentricity, semi-major axis and mass ratio. This may help to identify the link between $\bar{e}_d$ and $e_f$.

\begin{acknowledgements}
S.L and Z.M.L are supported by the STFC. P.J.C is grateful to NERC Grant NE/K004778/1. SJP is supported by a Royal Society University Research Fellowship. The authors acknowledge the University of Bristol Advanced Computing Research Centre facilities (https://www.acrc.bris.ac.uk/), and the use of DICE which were both used to carry out this work. The authors would also like to thank the referee S. Lubow for his insightful report.
\end{acknowledgements}

%-------------------------------------------------------------------

\end{document}